\pdfoutput=1

\documentclass[%
 reprint,
 floatfix,
 widetext,
 amsmath,amssymb,
 aps,
]{revtex4-2}

\usepackage{float}

\usepackage[strict]{changepage}
\usepackage[title]{appendix}
\usepackage{xcolor}
\usepackage{graphicx}
\usepackage{dcolumn}
\usepackage{bm}

\usepackage[utf8]{inputenc}
\usepackage{physics}
\usepackage{amsfonts}
\usepackage{amsmath}
\usepackage{xcolor}
\usepackage{array}
\newcolumntype{M}[1]{>{\centering\arraybackslash}m{#1}}
\usepackage{float}

\begin{document}

\preprint{APS/123-QED}

\title{Conductive metal oxide and hafnium oxide bilayer ReRAM: an ab initio study}

\author{Antoine Honet
}
\altaffiliation{Present address: Spectroscopy, Quantum Chemistry and Atmospheric Remote Sensing (SQUARES), Université libre de Bruxelles (ULB), Brussels, Belgium }
\affiliation{%
NanoComputing Research Lab, Department of Electrical Engineering, Eindhoven University of Technology, Eindhoven 5612 AP, The Netherlands
}%

\author{Aida Todri-Sanial}
\affiliation{%
NanoComputing Research Lab, Department of Electrical Engineering, Eindhoven University of Technology, Eindhoven 5612 AP, The Netherlands
}%

\date{\today}

\begin{abstract}

We perform generalized gradient approximation (GGA) simulations of interfaces between two Conductive Metal-Oxides (CMO, namely TaO and TiO) and cubic hafnium oxide ($HfO_2$) in the context of bilayer Resistive Random Access Memory (ReRAM) devices. We simulate filamentary conduction in $HfO_2$ by creating an atomically thin O atom vacancy path inside $HfO_2$. We show that this atomically thin filament leads to a great reduction of the resistance of the structures. Moreover, we explore the possibility of the influence of O excess inside the CMO on the global resistance of the device and confirm the induced modulation. We also shed the light on two possible causes for the observed increas in the resistance when O atoms are inserted inside the CMO. Eventually, we push forward key differences between devices with TaO and TiO as CMO. We show that structures with TaO are more stable in general and lead to a behaviour implying only low and high resistance (two well separated levels) while structures with TiO allows for intermediate resistances. 
 
\begin{description}
\item[Keywords]
Bilayer ReRAM, Hafnium oxide, conductive metal oxide, density functional theory, high and low-resistive states
\end{description}
\end{abstract}

\maketitle


\author{Antoine Honet, Aida Todri-Sanial}

\section{Introduction}

The increasing consumption of computational resources as well as increasing amount of data and use of artificial intelligence lead to the need of developing energy-efficient new devices and new computing paradigm. Neuromorphic in-memory computing is one of the potential paradigms and needs some efforts to develop appropriate and energy-efficient new devices.

Resistive Random Access Memory (ReRAM) are devices exhibiting several levels of resistance, with at least two levels: a low-resistance state (LRS) and a high-resistance state (HRS). Non-volative memory are key building blocks for new neuromorphic in-memory computing paradigm~\cite{begon-lours_analog_2021, carapezzi_role_2022, stecconi_filamentary_2022, falcone_physical_2023}.

In this view, $HfO_2$ has been widely studied both experimentally and theoretically for implementing such ReRAM devices. It is now widely accepted that the changes in resistance in pure $HfO_2$ ReRAM device are due to the formation of conductive filament made of oxygen vacancies~\cite{cartoixa_transport_2012, kim_physical_2013, kim_comprehensive_2014, padovani_microscopic_2015, celano_imaging_2015, dirkmann_filament_2018, xu_kinetic_2020, zeumault_tcad_2021, falcone_physical_2023}.

To the best of our knowledge, most of \textit{ab initio} studies focus on modelling this filamentary phenomenon either considering one to few O vacancies not arranged in a filament geometry~\cite{capron_migration_2007, sharia_extended_2009, ohara_assessing_2014, clima_first-principles_2016, traore_hfo_2018, gao_mechanisms_2019} or interfaces between $HfO_2$ and O defective material such as $HfO$, $Hf_2O_3$ or $Hf$~\cite{ohara_assessing_2014, wu_filament--dielectric_2017, padilha_structure_2018}. These studies aim at deriving quantities as energy barriers or Schottky barriers that can be used to feed larger-scale approaches such as Kinetic Monte Carlo or TCAD.

Exception is made for the work~\cite{cartoixa_transport_2012} were filament-shaped O vacancies in $HfO_2$ were considered. In particular, the authors studied transmission properties of atomically thick filament for monoclinic and amorphous $HfO_2$ sandwiched between metallic electrodes. It was shown that for the monoclinic case, already single atomic vacancy filaments led to significant increase in the transmission while larger filament region was needed for the amorphous case. As in~\cite{wu_filament--dielectric_2017}, we considered cubic $HfO_2$ as it usually results in more easily generated interfaces, with less strain and smaller supercell than the monoclinic phase~\cite{luo_combined_2008, wu_filament--dielectric_2017}, which is known to be the most stable crystallographic phase~\cite{luo_combined_2008, laudadio_phase_2022}. At the same time, the cubic phase of $HfO_2$ is well studied for its ferroelectricity behavior and can be stabilized either through doping (as Yttrium) or annealing/heating procedure~\cite{elshanshoury_polymorphic_1970, lim_dielectric_2002, luo_combined_2008,muller_ferroelectricity_2011}.

On the experimental part, several studies left the common single-layer $HfO_2$ ReRAM (sandwiched between two electrodes) to go for the investigation of bilayer devices made of a conductive metal-oxide (CMO) layer and a $HFO_2$ layer~\cite{hardtdegen_improved_2018, stecconi_filamentary_2022,  falcone_physical_2023}, or with other material replacing $HfO_2$ for the formation of the filament region as $Ta_2O_5$~\cite{kim_comprehensive_2014}. These experimental studies are often accompanied by theoretical models using macroscopic electromagnetism or drift-diffusion methods. It was suggested in~\cite{stecconi_filamentary_2022, falcone_physical_2023} that, combined with the traditional filament conduction mechanism in the $HfO_2$ layer, a mechanism of oxidation inside the CMO layer could lead the resistance switching mechanism observed in these devices. These oxidized regions would arise from O migration from the conductive filament region and acts as a bottleneck in the conductivity such that they are associated with HRS. There is, to the best of our knowledge, no \textit{ab initio} studies on these new bilayer devices were performed up to now and the microscopic details implied in the switching mechanism are yet to be unraveled. They are of primary importance since their understanding could lead to more efficient tuning and engineering of the resistance levels in ReRAM devices.

In this article, we performed density functional theory (DFT) \textit{ab initio} simulations of the interface between two CMO (TaO and TiO) and $HfO_2$. With the aim of describing different possible resistance levels, we model atomically thin O-vacancy filament inside the $HfO_2$ material and we also model O excess inside the CMO layer using different geometrical starting point for the added O atoms. We will show that our model can reproduce the strong decrease in resistance associated with the creation of a filament but also that the resistance can be modulated \textit{via} the location of the O excess atoms in the CMO layer. 

Furthermore, we highlight fundamental differences associated with the choice of the CMO material, either TaO or TiO. In particular, we show that the considered structures are more stable when TaO is employed. TaO exhibit resistances that lay in two different regime (high and low) while it turns out that TiO interfaces result in more spread resistances.

To sum up, in this paper we:
\begin{itemize}
    \item a first DFT study of CMO-$HFO_2$ bilayers
    \item investigate the resistance's drop due to filament formation in the $HfO_2$ layer
    \item highlight the effect of the considered CMO, comparing TaO and TiO
    \item investigate the creation of oxygen excess in the CMO layer, potentially due to migration from the $HFO_2$ layer
    \item highlight different interpretations of the change in resistance studying electron localization function.
\end{itemize}

The following of the paper is organized as follow: we first introduce at section~\ref{sec:methods} the methods used for our study as well as the technical description of the considered interfaces, the modelling of the filament and location of O excess atoms inside CMO. Then, we expose in section~\ref{sec:results} the obtained results focusing on the stability and the change in resistance of the studied structures. We end this article with a discussion and a conclusion, containing perspectives for further research.

\section{Models and methods}
\label{sec:methods}

\subsection{Numerical tools}
\label{sec:DFT_tools}

Throughout this paper, we used the DFT software QuantumATK V-2023.09~\cite{QATK_2023}. We performed our computations using the LCAO solver, combined with generalized gradient approximation GGA (PBE) functional~\cite{perdew_generalized_1996} and PseudoDojo pseudopotentials~\cite{van_setten_pseudodojo_2018} in their medium implementation. An energy cut off of $125 Ha$ and a k-point 6 x 6 x 50 Monkhorst-Pack~\cite{monkhorst_special_1976, pack_special_1977} grid for the $HfO_2$ electrodes and the central regions and a k-point Monkhorst-Pack grid of 12 x 12 x 50 for the TaO and TiO electrodes were used, while we fixed the broadening at $1000 K$ for the Fermi-Dirac distribution.

We used the two-probe device setup~\cite{stradi_general_2016} implemented in QuantumATK and the interfaces were built using the supercell builder implemented in the NanoLab GUI of QuantumATK~\cite{stradi_method_2017}.

Transport properties were computed using the Non-Equilibrium Green's Function formalism and resistances were extracted from the Landauer formula~\cite{brandbyge_density-functional_2002}. The k-point grid for transport properties was a 8x8 Monkhorst-Pack grid and we used an energy resolution of 0.04 eV in the interval [-4 eV, 4 eV]. This resolution was found to be sufficient for convergence as shown in the SI, as well as k points convergence studies.

\subsection{Interfaces}

\begin{figure}[]
\centering
    \includegraphics[width=8cm]{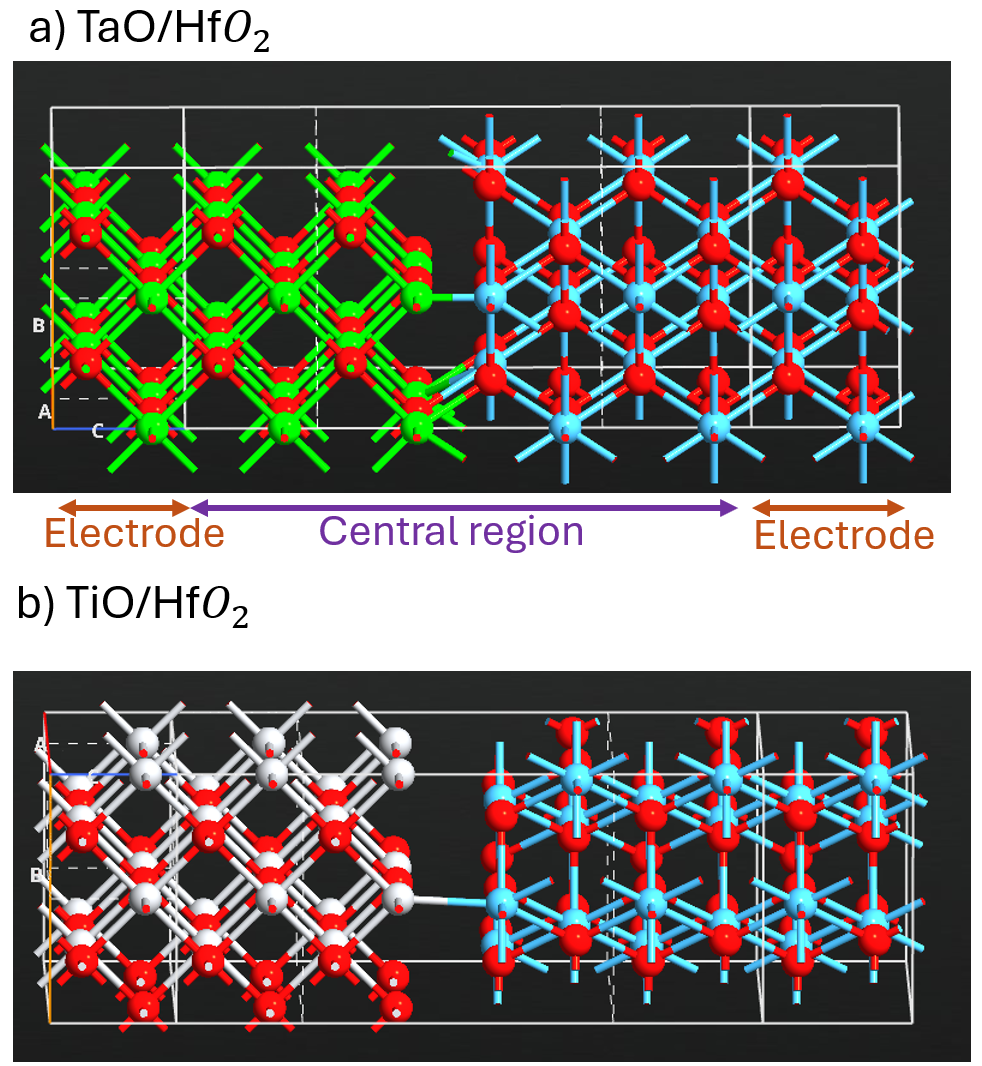}
\caption{Illustration of the device configurations investigated, for TaO/$HfO_2$ (a) and TiO/$HfO_2$ (b). Devices are made of a central region and two electrodes, repeated periodically across the transport direction. Ta atoms are represented in green, O atoms in red, Hf atoms in blue and Ti atoms in grey.}
\label{fig:TaO_and_TiO_devices}
\end{figure}

In our work, we considered two distinct interfaces, first between cubic $TaO$ and cubic $HfO_2$ and second between cubic $TiO$ and cubic $HfO_2$. All the bulk structures were first relaxed before building the interfaces, starting from the QuantumATK database data. The resulting lattice parameters are found in Table~\ref{tab:lattice_parameters}.

\begin{table}
    \centering
    \begin{tabular}{|M{30mm}|M{10mm}|M{10mm}|M{10mm}|}
        \hline
    	&  \multicolumn{3}{c}{Lattice parameters ($\AA$)} \vline \\
    	\hline
        & TaO & TiO & $HfO_2$  \\
        \hline
        From database & 4.422 & 4.177 &  5.115  \\
        Optimized & 4.508 & 4.294 & 5.077  \\

        \hline
    \end{tabular}
    \caption{Lattice parameters for cubic unit cells of $TaO$, $TiO$ and $HfO_2$, both from the QuantumATK database and after optimization.}
    \label{tab:lattice_parameters}
\end{table}

Two interfaces between $HfO_2$ [110] surface and the two conductive metal oxides [110] surfaces were built using the optimized geometries, looking for surfaces with a balance between low strain and small size of the supercell. The $TaO$ and $TiO$ lattice parameters being very close, our two chosen supercells are similar. The parameters of the supercells are given in Table~\ref{tab:parameters_supercells}. After using the module, we made small modifications for the TiO/$HfO_2$ interface, to ensure that a complete plane of 4 Ti atoms and 4 O atoms was present at the interface. Each interface contains a total of 68 atoms (16 Ta/Ti, 12 Hf, and 40 O). From the interfaces, device two-probe configurations were built, which are shown at Fig.~\ref{fig:TaO_and_TiO_devices}. They consist of a central region made of the 68-atom interface and two electrodes (TaO/TiO for left electrode and $HfO_2$ for right electrode) that are repeated periodically across the transport direction (the direction crossing the interface).

\begin{table}
    \centering
    \begin{tabular}{|M{30mm}|M{15mm}|M{20mm}|M{20mm}|}
    	\hline
        Interface& Area ($\AA^2$) & Rotation between surfaces (°) & Mean absolute strain ($ \%$)  \\
        \hline
        $TaO[110]/HfO_2[110]$ & 57.48 & 54.74 &  1.68  \\
        $TiO[110]/HfO_2[110]$ & 52.15 & 125.26 & 1.56  \\
        \hline
    \end{tabular}
    \caption{Interfaces supercells' parameters for the two constructed interfaces between $HfO_2$ and the two CMO (TaO and TiO).}
    \label{tab:parameters_supercells}
\end{table}

\subsection{Filaments modelling}

\begin{figure}[]
\centering
    \includegraphics[width=8cm]{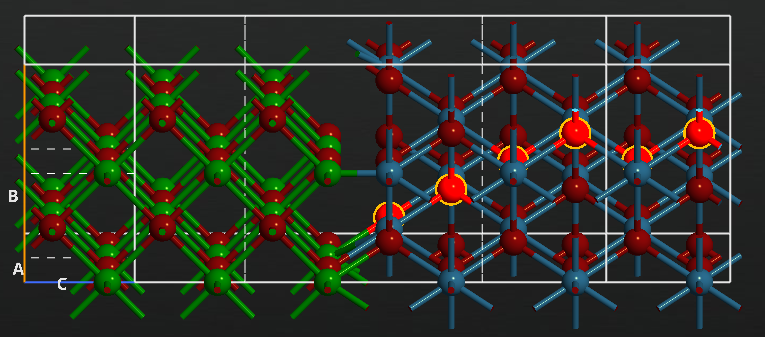}
\caption{Illustration of the O vacancies considered to modelled a filament in the $HfO_2$ part of the TaO/$HfO_2$ device. Ta atoms are represented in green, O atoms in red, Hf atoms in blue and the highlighted red atoms are the O atoms removed to model a filament.}
\label{fig:TaO_cHfO2_filament}
\end{figure}

Typical $HfO_2$ ReRAM devices are understood to present a filament of O-defective region enabling transport in the $HfO_2$ layer~\cite{cartoixa_transport_2012, kim_physical_2013, kim_comprehensive_2014, padovani_microscopic_2015, celano_imaging_2015, dirkmann_filament_2018, xu_kinetic_2020, zeumault_tcad_2021, falcone_physical_2023}. We modelled this filament in the right part of the device configurations of Fig.~\ref{fig:TaO_and_TiO_devices} by removing one O atom in each $HfO_2$ layer of the central region, forming a diagonal filament of O vacancies. This is illustrated at Fig.~\ref{fig:TaO_cHfO2_filament} for TaO/$HfO_2$ device and done in a similar way for the TiO/$HfO_2$ case. In the following of the paper, we refer to these configurations as "F".

\subsection{Oxygen excess inside the CMO layer}

A current interpretation of bilayer CMO/$HfO_2$ ReRAM mechanism is the creation of an oxygen-rich region inside the CMO that will act as a bottleneck for the conductivity and increase the resistance of the device~\cite{stecconi_filamentary_2022, falcone_physical_2023}. We considered several initial configurations to model this oxygen-rich region. In each case, we inserted four O atoms either in the closest CMO atomic layer to the interface, in the second atomic layer or in between. For each distance to the interface, we placed the O atoms in three different place inside the considered plane. This procedure leads to 9 different configurations for each of the CMO (TaO and TiO).

\begin{figure}[]
\centering
    \includegraphics[width=8cm]{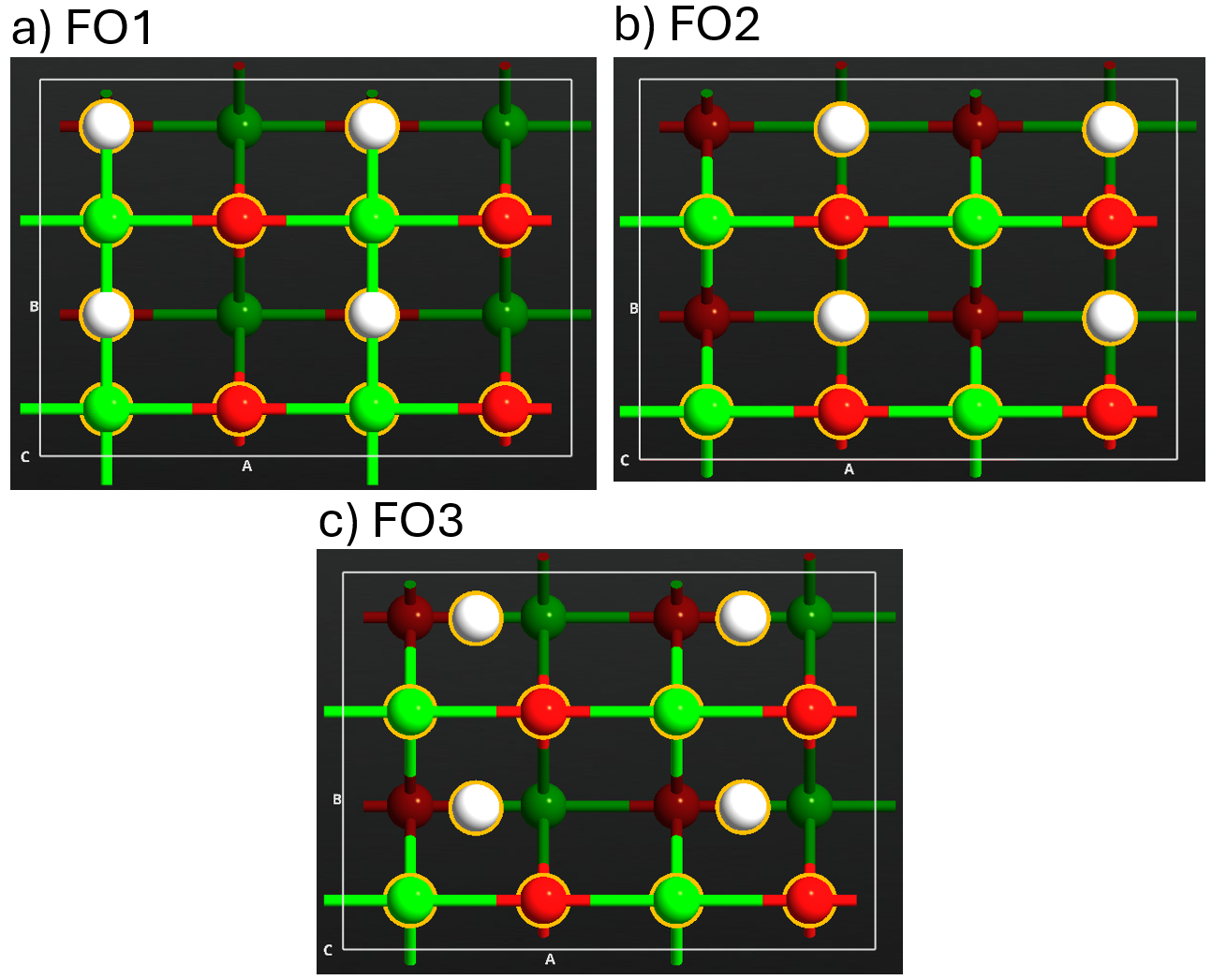}
\caption{Illustration of the added O in the TaO first layer for the TaO/$HfO_20$ device, for the three different configurations, in a front view. Ta atoms are depicted in green, original O atoms in red and added O atoms in white. The highlighted atoms belong to the first atomic layer from the interface while the atoms while the shadowed atoms belong to the second atomic plane from the interface. In FO1 (a), four O atoms are added between Ta atoms of the first atomic layer, and in front of the O atoms of the second atomic layer; in FO2 (b), four O atoms are added between O atoms of the first atomic layer and in front of the Ta atoms of the second atomic layer and in FO3 (c), four O atoms are added in the center of rectangles made of two Ta and two O atoms in the first atomic plane.}
\label{fig:FO1_2_3_front_views}
\end{figure}

For the three configurations with O atoms inserted inside the first CMO layer, we considered adding them in the form of a geometrical rectangle, either between O or metal atoms in the same layer or in the middle of a rectangle composed of two O and two metal atoms. We named these configurations FO1, FO2 and FO3 and illustrate the them for TaO/$HfO_2$ at Fig.~\ref{fig:FO1_2_3_front_views}.

Secondly, the same configurations as FO1, FO2 and FO3 are replicated between the two first atomic layer of TaO (equidistantly) and resulting in FO4, FO5 and FO6 configurations, respectively. Eventually, the three last configurations considered are in the second TaO atomic layer includes four O atoms in the same flavour as in the first atomic layer: between Ta atoms of the second layer and in front of O atoms of the first layer (FO7), between O atoms of the second layer and in front of Ta atoms of the first layer (FO8) and at the center of rectangles formed by two Ta atoms and two O atoms in the second layer plane (FO9).

\subsection{Optimized structures and energies}

We optimized our structures using a force tolerance of $0.05 \hspace{0.1cm} eV \hspace{0.1cm} {\AA}^{-1}$, leading to a total energy converged with smaller tolerance than $4 \vdot 10^{-5} \hspace{0.1cm} eV/Atom$. The formation energies of the central regions ($E_{\mathrm{form}}$) are defined as:

\begin{equation}
E_{\mathrm{form}}  =\begin{cases} E_{\mathrm{F}} + 2 E_{0_2} - E_{\mathrm{No Vac}} & \mathrm{(pure\hspace{0.1cm} filaments)} \\
                     E_{\mathrm{FOi}} - E_{\mathrm{No Vac}} &  \mathrm{(filaments \hspace{0.1cm} \& \hspace{0.1cm} O \hspace{0.1cm}excess)} 
       \end{cases}
\end{equation}
where $E_{\mathrm{F}}$, $ E_{\mathrm{FOi}}$ and $E_{\mathrm{No Vac}}$ are respectively the total energies of the geometry with a filament, with a filament and an oxygen excess in the CMO material and without any modification from the CMO/$HfO_2$ bulk interface. $E_{0_2}$ is the total energy of the di-oxygen molecule and is present in the formation energy of pure filament cases because there is a lack of four oxygen as compared to the non defective case. All the total energies are evaluated within the same DFT-GGA framework exposed at section~\ref{sec:DFT_tools}, after geometry optimization.

\section{Results}
\label{sec:results}

\begin{figure*}[]
\begin{centering}
    \includegraphics[width=\textwidth]{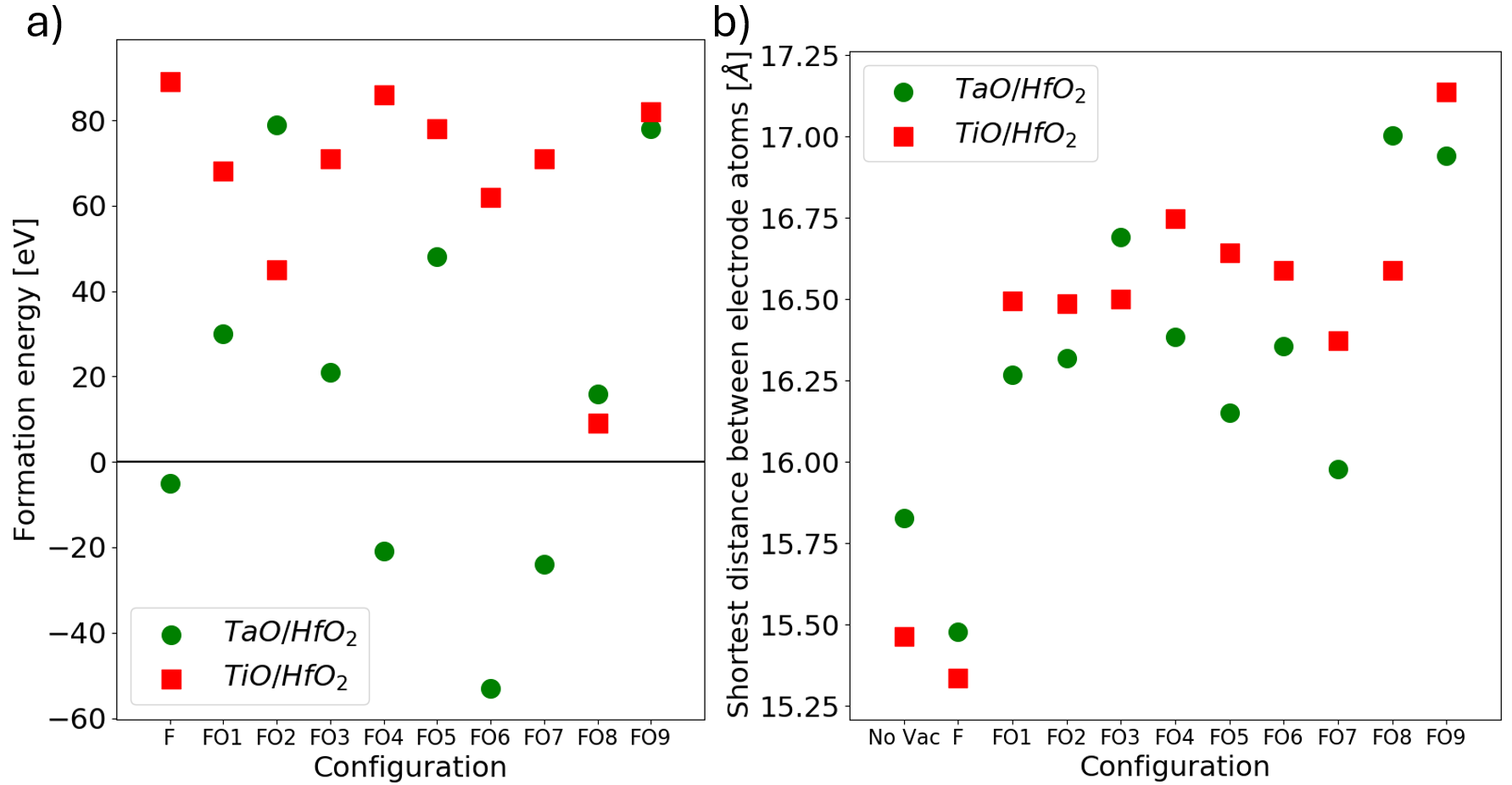}
\caption{a) Formation energies of the ten configurations considered, \textit{i.e.}, filament structures (F) and filament structures with O excess in the CMO material (FOi). The results are shown for TaO (green circles) and TiO (red squares) as CMO material. b) The shortest distance between atoms in the left and right for the different configurations and the two CMO. The symbols and colors refer to the same CMO as in a).}
\label{fig:E_form_Electrode_distance}
\end{centering}
\end{figure*}

For both CMO, the 11 devices (non defective, filament and 9 filament + O excess) were performed geometry optimization and computed the formation energies. Fig.~\ref{fig:E_form_Electrode_distance}~a) shows the formation energies of the ten structures for both TaO and TiO considered as CMO.

First, we see for the formation energies at Fig.~\ref{fig:E_form_Electrode_distance}~a) that TiO as CMO only exhibit one configuration with low formation energy, \textit{i.e.}, the FO8 configuration with $E_{\mathrm{form}} \simeq 9 \hspace{0.1cm} eV$ while there are 6 configurations with TaO with $E_{\mathrm{form}}$ around or below $20 \hspace{0.1 cm} eV$. Moreover, four configurations with TaO exhibit a negative formation energy meaning that the total energy is lower than the non defective interface case. What's more, we see that the most stable FO configurations tend to be either between the first and second atomic layer of TaO or within the second atomic layer, especially not within the first atomic layer. The configurations with O a bit away from the interface therefore seem to be more stable than with O at the interface.

Importantly, for TiO as CMO, the pure filament configuration exhibit the highest formation energy, making it highly unstable and the inclusion of O excess in CMO favored in all cases, which is a notable difference with devices implying TaO as CMO. The shortest distance between atoms of left and right electrodes are reported at Fig.~\ref{fig:E_form_Electrode_distance}~b). In both cases, the inclusion of the filament leads to smaller central region then for the non-defective case, even though the effect is more important when TaO is used. Then, including of O atoms in the CMO increase the size of the central region, that turns out to be larger than the non-defective central region for each FOi configuration. Inclusion inside the first layer and between the first and second layers (FO 1-6) leads to intermediate length while some of the configurations including the O atoms in the second CMO atomic layer increase the central region length even more (see FO8 and FO9 for TaO/$HfO_2$ and FO9 for TiO/$HfO_2$ at Fig.~\ref{fig:E_form_Electrode_distance}~b)). 

\begin{figure}[]
\centering
    \includegraphics[width=8cm]{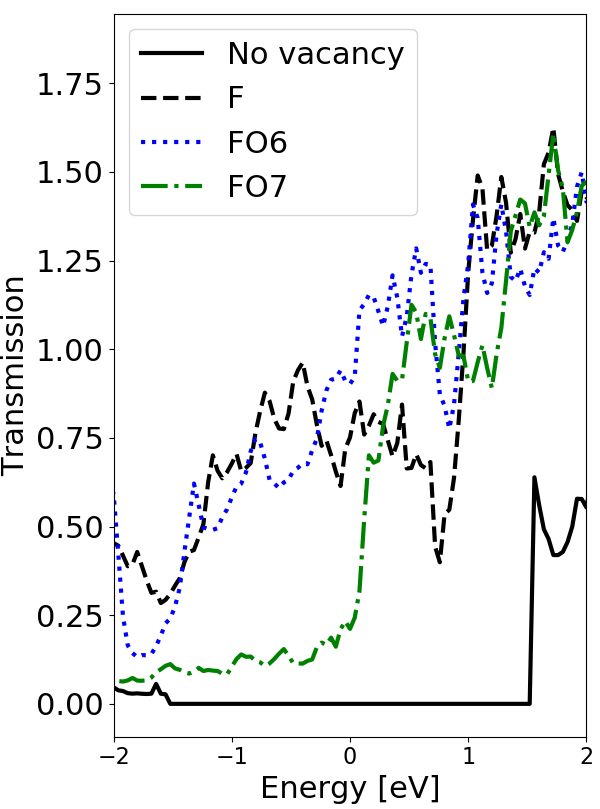}
\caption{Transmission spectrum between $-2$ and $2 \hspace{0.1cm} eV$ for the four configurations of the TaO/$HfO_2$ device: the non-defective one (solid black line), the one with filaments inclusion (F, dashed black line) and the two lowest-energy filaments + O excess configurations, \textit{i.e.}, FO6 (blue dotted line) and FO7 (green dashdoted line).}
\label{fig:transmission1}
\end{figure}

We focus now on the transmission spectra features. In particular, we show on Fig.~\ref{fig:transmission1} the transmission spectra of the non defective, the pure filament as well as FO6 and FO7 configurations for the TaO/$HfO_2$ device, which are the two filament + O excess configurations with the lowest formation energy (see Fig.\ref{fig:transmission1}).

First, we see that for the non defective case, the transmission spectrum present a zero-transmission feature around the Fermi level. Transmission starts is non zero for energies larger then $\simeq 1.55 \hspace{0.1 cm} eV$ in absolute values. Including filaments in $HfO_2$ leads to a great increase in the transmission spectrum, especially at the Fermi level. The increasing transmission of monoclinic $HfO_2$ found in~\cite{cartoixa_transport_2012} for single-vacancy filaments is then also observed in the cubic phase. Then, including excess O in the CMO material can either slightly increase the transmission of the transmission around Fermi level (see FO6 at Fig.~\ref{fig:transmission1}) or decrease it (FO7 on the same figure).

\begin{figure}[t]
\centering
    \includegraphics[width=8cm]{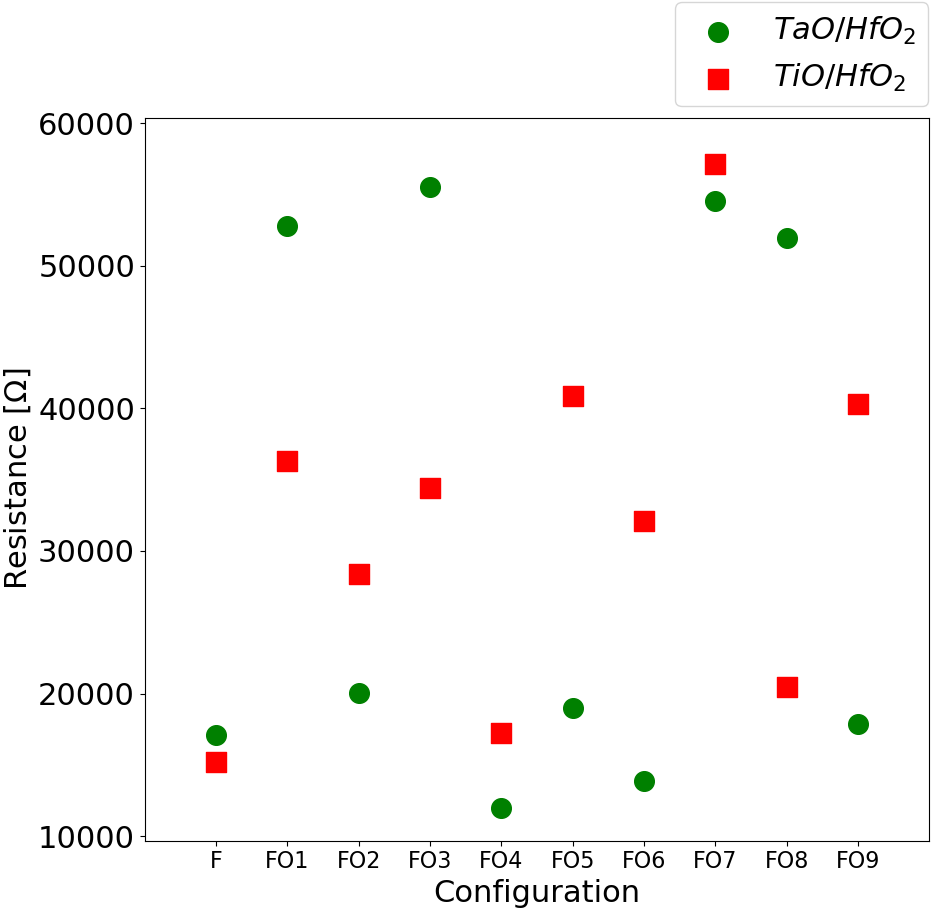}
\caption{Resistance of the filament (F) and the nine filament + O excess (FOi) device configurations for both TaO/$HfO_2$ (green circles) and TiO/$HfO_2$ (red squares).}
\label{fig:resistances}
\end{figure}

We computed the resistance of all the configurations for both materials from the Landauer formalism implemented in QuantumATK. First, when going from a non-defective device to a device presenting a O-vacancy filament, the resistance drops down radically from $1.55 \vdot 10^{30} \hspace{0.1 cm} \Omega$ for TaO/$HfO_2$ (resp. $1.87 \vdot 10^{32} \hspace{0.1 cm} \Omega$ for TiO/$HfO_2$) to $1.71 \vdot 10^{4} \hspace{0.1 cm} \Omega$ (resp. $1.52 \vdot 10^{4} \hspace{0.1 cm} \Omega$). The resistance for the filament and filament + O excess in CMO are shown at Fig.~\ref{fig:resistances}.

From Fig.~\ref{fig:resistances}, we see that the inclusion of O excess in TaO as CMO leads to either a high resistance (although still a lot smaller than the non-defective interface) of $50-60 \hspace{0.1cm} k\Omega$ or a small resistance of the order of magnitude of the one without O excess ($10-20 \hspace{0.1cm} k\Omega$). These values are in agreement with the LRS and HRS resistances found experimentally in~\cite{stecconi_filamentary_2022}. This behaviour of either small or high resistance relates to a kind of binary behaviour. What's more, the two most stable configurations (FO6 and FO7, see Fig.~\ref{fig:E_form_Electrode_distance}~a)) exhibit resistances that are in the different regimes. 

On the other hand, when looking at TiO as CMO on Fig.~\ref{fig:resistances}, the spectrum of possible resistance is more contiunous, with relatively small resistances for three configurations (F, FO4 and FO8), five intermediate resistances and a high resistance (FO7). The most stable structure (FO8, see Fig.~\ref{fig:E_form_Electrode_distance}~a)) exhibit a relatively small resistance and the second most stable (FO2) and intermediate one, while the other configurations are closer in energy (see Fig.~\ref{fig:E_form_Electrode_distance}~a)). The range and distribution in resistance for TiO devices observed at Fig.~\ref{fig:resistances} could lead to multilevel resistance ReRAM devices, in strong contrast with TaO based CMO devices where only high and low resistances are observed. Note however that the maximum ratio between HRS and LRS are of the same order of magnitude for both TaO/$HfO_2$ (4.63) and TiO/$HfO_2$ (3.76). 

\begin{figure}[]
\centering
    \includegraphics[width=8cm]{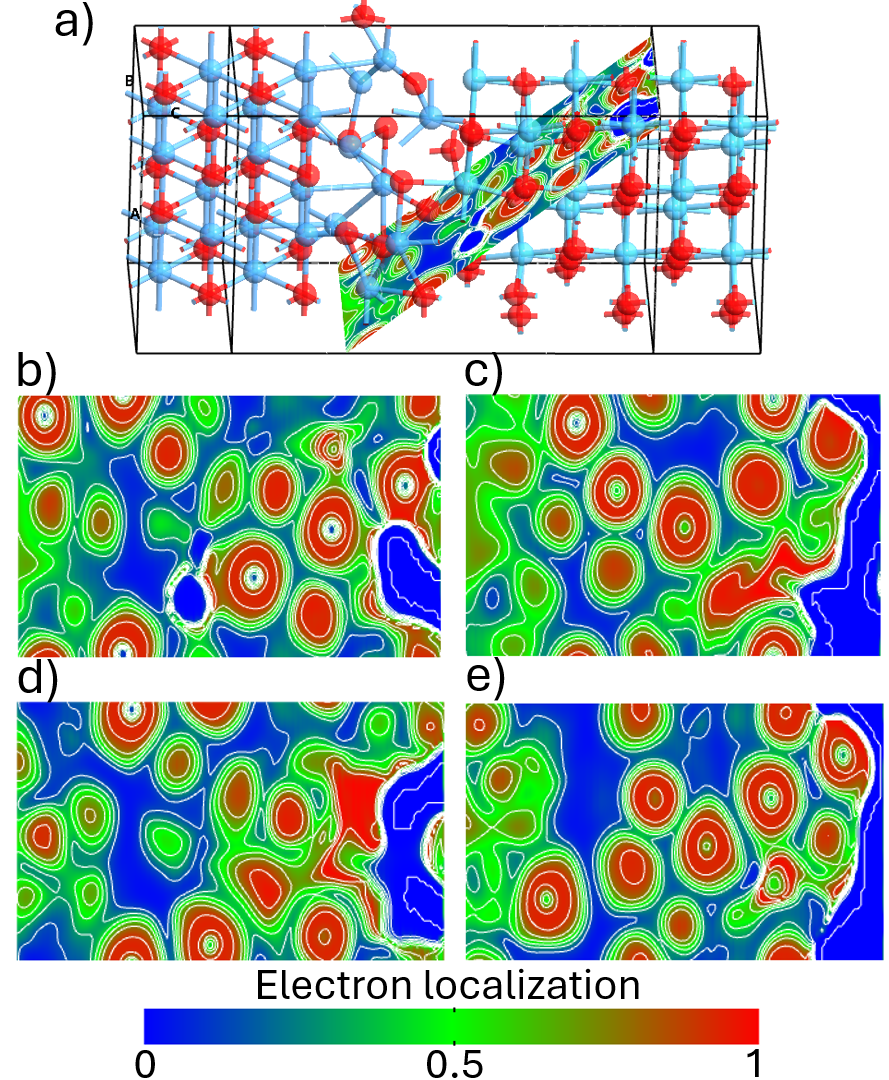}
\caption{a) Illustration of the cut planes shown in b), c), d) and e) for the optimized non-defective TaO/$HfO_2$ interface. The cut plane is located at the proximity of the four removed atoms for the filament modeling. In b), c), d) and e), we show a top view of this cut plane for the non-defective, F, FO6 and FO7 configurations of the TaO/$HfO_2$ interface.}
\label{fig:ELF_top_views}
\end{figure}

We now focus on the electron localization function (ELF) feature to better describe the changes between our simulated devices. Let us recall that the ELF ranges between 1 and 0, that an ELF value close to 1 reflects strongly localized electrons while an ELF value close to 0.5 reflects delocalized electron and more metallic bonds. ELF cut planes in the vicinity of the atoms removed for the filament are shown at Fig.~\ref{fig:ELF_top_views} b), c), d) and e) for non-defective, F, FO6 and FO7 configurations of the TaO/$HfO_2$ interface respectively with top views. The spatial localization of the plane is shown for illustration at Fig.~\ref{fig:ELF_top_views} a) for the non-defective case. 

We see, comparing Figs.~\ref{fig:ELF_top_views} b) and c) that the inclusion of the filament creates a connecting path of ELF $\simeq 0.5$ in the plane where the O atoms are removed, understood as a responsible feature of the decrease of the resistance. Then, the inclusion of O excess in TaO can preserve (even extend) this connecting path as for the FO6 structure, see Fig.~\ref{fig:ELF_top_views} d). This correlates well with the low resistance of this structure reported at Fig.~\ref{fig:resistances}. However, the inclusion of O excess atoms in TaO can also spoils out the connecting path inside the $HfO_2$ layer as shown on Fig.~\ref{fig:ELF_top_views} e) for FO7. This, again well correlates with Fig.~\ref{fig:resistances} that predicts a high resistance for the FO7 configuration. 

\begin{figure}[]
\centering
    \includegraphics[width=8cm]{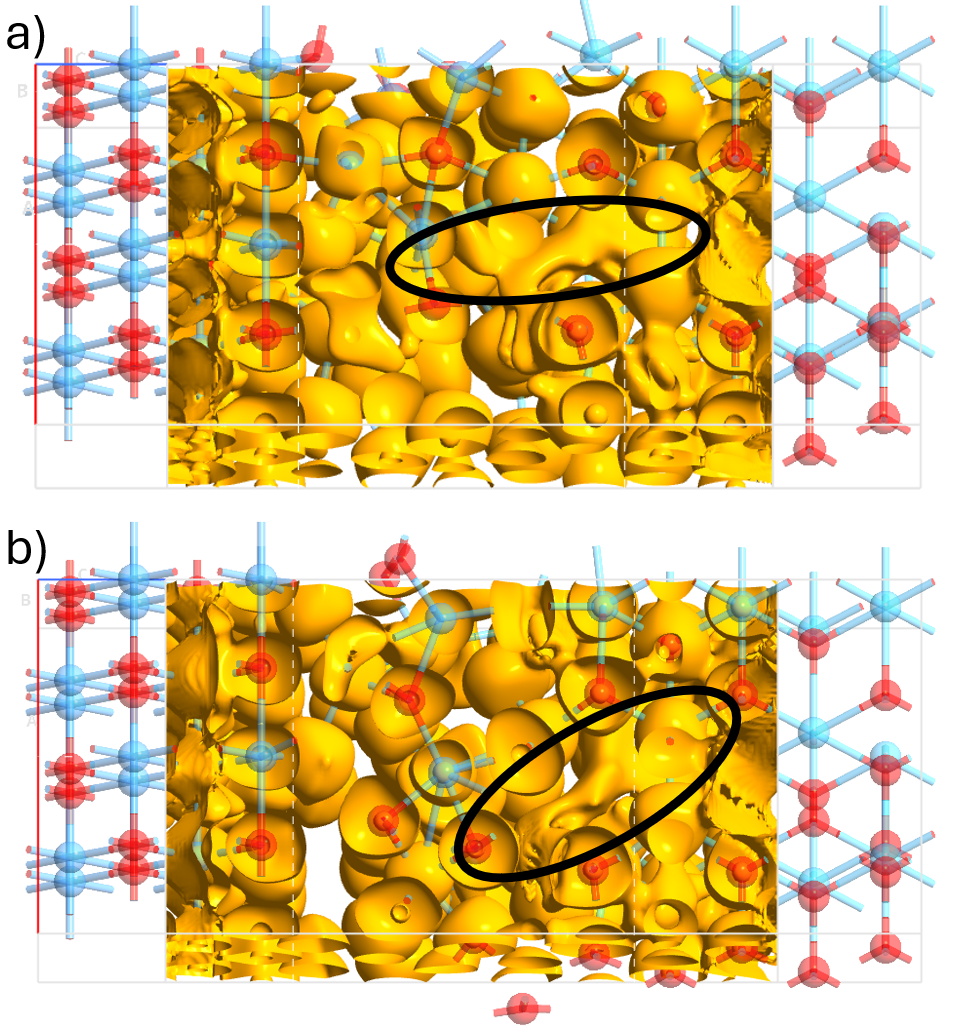}
\caption{Isosurface of ELF at the value 0.65 for the optimized FO4 (a) and FO8 (b) structures for TaO/$HfO_2$ interface. The oval shapes on the figures highlight the regions where an ELF path in $HfO_2$ is present and the connection at the interface.}
\label{fig:ELF_iso}
\end{figure}

In addition to the destruction of the ELF path in the $HfO_2$, we identified another factor of resistance increase, following the inclusion of O excess atoms in a CMO layer. In Fig.~\ref{fig:ELF_iso}, we compare ELF isosurface at 0.65 for FO4 and FO8, presenting respectively low and high resistance (see Fig.~\ref{fig:resistances}). In this figure, we can see that an intermediate ELF path is present in both structure inside the $HfO_2$ material. However, there is a major difference between these two structure ELF features. In the FO4 configuration (Fig.~\ref{fig:ELF_iso}~a)) the path along the removed O atoms in $HfO_2$ connects at the interface to a another intermediate ELD region inside the TaO material. In the FO8 configuration (Fig.~\ref{fig:ELF_iso}~b)), the intermediate region in $HfO_2$ does not connect to a similar region at the interface and is bounded to $HfO_2$, resulting in breaking the metallic path.

\section{Discussion, conclusion and perspectives}

\subsection{Discussion and conclusion}

We studied in this paper interfaces between $HfO_2$ and CMO (TaO and TiO) in the context of bilayer ReRAM devices, using \textit{ab initio} techniques. Bilayer ReRAM are new devices, currently foreseen to be of great interest to be used for in-memory neuromorphic computing, allowing for more energy-efficient computations.

Since pure $HfO_2$ ReRAM devices are often described as filamentary devices and bilayer CMO/$HfO_2$ were also suggested recently to be filamentary based devices~\cite{stecconi_filamentary_2022, falcone_physical_2023}, we modelled the change in resistance of the device using filament O vacancies in $HfO_2$ and O excess inside the CMO, as suggested in the two same references.

In total, we simulated 22 structures, \textit{i.e.} one non-defective, one containing pure filament and nine different filament + O excess configurations for each of the two CMO.

For all the 22 optimized structures, the formation energies and central region length were computed and reported. We showed that the inclusion of filament and O excess lead to more stable structures with TaO as CMO than with TiO. In addition, the most stable configurations including O excess in the CMO are always for the configurations for which we incorporated O atoms beyond the first atomic layer of CMO. On the other hand, no direct link with the central region length was evidenced from our work. 

In the context of ReRAM, we focused on the transmission spectra and computed resistances of the devices. We showed that the inclusion of a filament of O vacancies inside $HfO_2$ using \textit{ab initio} technique can lead to a drastic decrease in the resistance of the device. We also highlighted that the inclusion of O excess atoms in the CMO layer can modulate the resistance and pushed forward two reasons for an increase in the resistance. Either the inclusion of the O excess atoms in the CMO can destroy the filament induced ELF $\simeq 0.5$ path or the O excess inclusion can make the filament not connecting to a proper ELF$\simeq 0.5$ region in the CMO at the interface.

With this \textit{ab initio} study, we give clues for physical grounds of an earlier adopted continuous model in~\cite{stecconi_filamentary_2022, falcone_physical_2023}.

From the comparison between TaO/$HfO_2$ and TiO/$HfO_2$, the behaviour difference in terms of two-level LRS/HRS (TaO) or more continuous resistance (TiO) suggests important technological implications, making TaO more suitable for two-level ReRAM which is also accentuated by the favored stability of the structures. However, TiO on the contrary could be investigated for multi-level ReRAM that are also of technological interest, probably at the price of more controlled defects engineering.

\subsection{Future perspectives}

In the future, we think that other \textit{ab initio} studies could be undertaken to investigate other crystalline structures (as monoclinic or amorphous that were studied for pure $HfO_2$ in~\cite{cartoixa_transport_2012}) and surface orientations or supercells, to see if our highlights can be generalized or if other behaviours can be observed. Moreover, in complement with the formation energies reported here, we believe that a more dynamical description for some of the configurations could lead to a better understanding of the formation mechanism of filament + O excess regions in bilayer ReRAM. Such studies would include nudge elastic band (NEB)~\cite{henkelman_climbing_2000, henkelman_improved_2000}  technique for computing intermediate states. The combination of the different O excess distribution and changing the concentration as well as the effect of the size (width) of the conductive filament are other open questions that are to be tackled in future studies. Eventually, studying finite-bias devices (IV characteristic) would be step to relate such \textit{ab initio} studies to experimental and technological implementations.


\section*{Acknowledgements}
This work is supported and funded under the framework of the Horizon Europe program, project PHASTRAC (\url{https://phastrac.eu}) with grant number 101092096.

\bibliographystyle{ieeetr}
\bibliography{bibliography}

\end{document}


\preprint{APS/123-QED}

\title{Conductive metal oxide and hafnium oxide bilayer ReRAM: an ab initio study}

\author{Antoine Honet
}

\affiliation{%
Department of Electrical Engineering, Eindhoven University of Technology, Eindhoven 5612 AP, Netherlands
}%

\author{Aida Todri-Sanial}
\affiliation{%
Department of Electrical Engineering, Eindhoven University of Technology, Eindhoven 5612 AP, Netherlands
}%

\date{\today}

\maketitle


\author{Antoine Honet, Aida Todri-Sanial}

We support in this supplementary information the results exposed in the main text by given insight into the convergence tests made for the choice of parameters used in the main text.

\section{Convergence in k points}

\begin{figure*}
\centering
    \includegraphics[width=16cm]{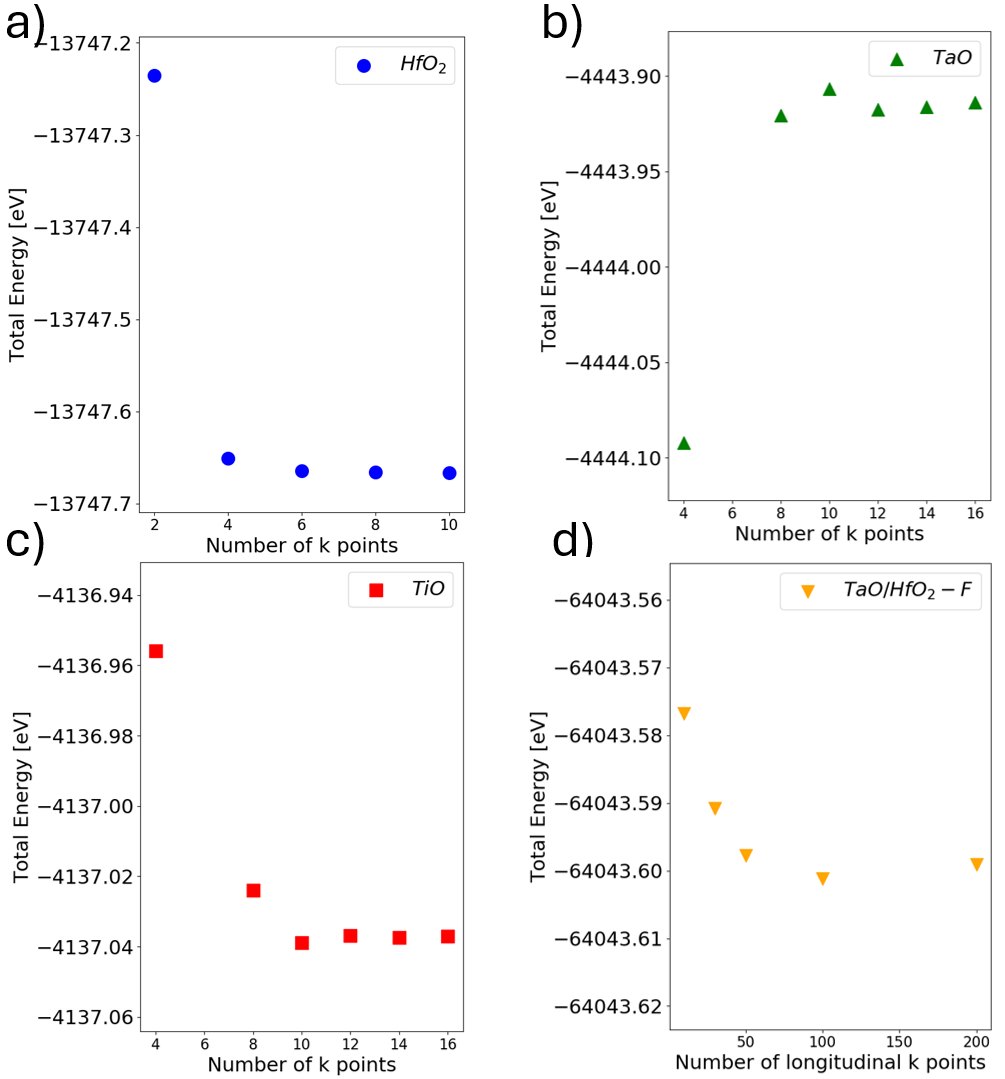}
\caption{Total energies as a function of k points in a Monkhorst-Pack k x k x k grid for the $HfO_2$ (a), TaO (b) and TiO (c) electrodes. d) shows the total energies as a function of the longitudinal number of k points (resulting in a Monkhorst-Pack 6 x 6 x k grid) for the filament configuration of the TaO/$HfO_2$ interface. }
\label{fig:conv_k_pts}
\end{figure*}

We used a 6x6 Monkhorst-Pack grid used for the non-transport directions in $HfO_2$ electrode and a 12 x 12 Monkhorst-Pack grid for the TaO and TiO electrodes. These lead to relative convergence of the total energies of the electrodes' bulk structures of $1.4 \vdot 10^{-7}$ (resp. $2.9 \vdot 10^{-7}$ and $1.3 \vdot 10^{-7}$) for $HfO_2$ (resp. TaO and TiO). The bandgap convergence for $HfO_2$ was within $2 \vdot 10^{-4} \hspace{0.1cm} eV$. The total energy convergence for these structures are shown at fig.~\ref{fig:conv_k_pts} a), b) and c).

For the device geometry, we used a 6 x 6 x 50 Monkhorst-Pack grid that lead to a relative convergence of less than $5.3\vdot 10^{-8}$ in the TaO/$HfO_2$ pure filament configuration. The evolution of total energy for this configuration is shown at fig.~\ref{fig:conv_k_pts} d). 

\section{Convergence of the transmission spectra}

In the main text, we specified that we used a k-point Monkhorst-Pack 8x8 grid and an energy resolution of 0.04 eV for the transmission spectra. The default energy resolution in QuantumATK is higher but in a more restricted energy interval since it is designed to cover the bias interval.

\begin{figure}
\centering
    \includegraphics[width=8cm]{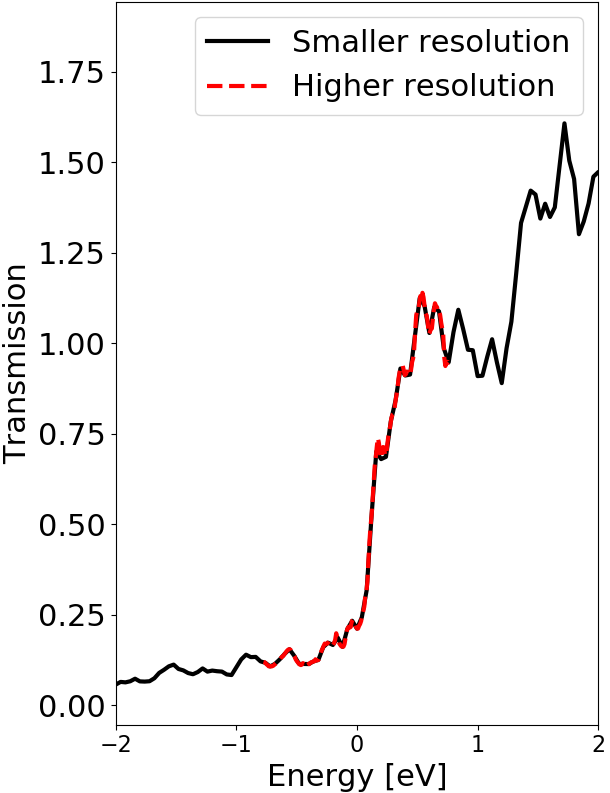}
\caption{Transmission spectrum for one of the lowest energy configurations of the TaO/$HfO_2$ device (FO7, see main text) showing two different resolutions. The smaller energy resolution is of 0.04 eV while the higher one is of 0.0082 eV. }
\label{fig:transmission_resolution}
\end{figure}

The smaller resolution considered 201 energy points in the energy interval [-4 eV, 4 eV] (energy resolution of 0.04 eV) while the higher resolution (default in QuantumATK) considered 180 energy points in the energy interval [-0.776 eV, 0.767 eV] (energy resolution of 0.0082 eV). The resulting transmission spectra for the two energy resolutions are shown at fig.\ref{fig:transmission_resolution} for the FO7 configuration of the TaO/$HfO_2$ interface. This figure illustrates that the energy resolution used in this paper is sufficient according to transmission spectrum convergence. 

\begin{figure}
\centering
    \includegraphics[width=8cm]{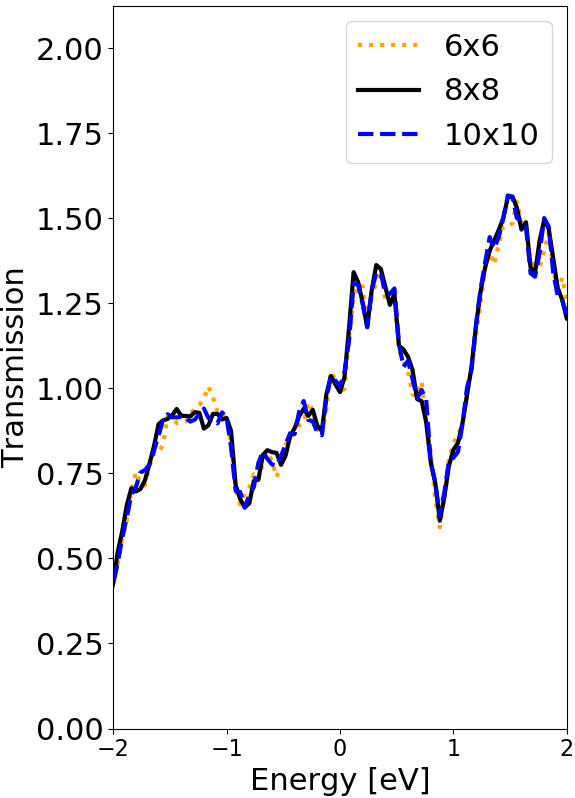}
\caption{Transmission spectrum for the pure filament non optimized configuration TaO/$HfO_2$ device (F, see main text) showing three different k points sampling for the transmission calculation: Monkhorst-Pack 6x6 (orange dotted curve), 8x8 (full black curve) and 10x10 (blue dashed curve).}
\label{fig:transmission_k_pts}
\end{figure}

We also report convergence tests for the k-point Monkhorst-Pack grid at fig.~\ref{fig:transmission_k_pts} for 6 x 6, 8 x 8 and 10 x 10 grids, for the non optimized TaO/$HfO_2$ pure filament configuration. We see that the 8 x 8 grid (used in the main text) leads to good transmission convergence.

\bibliographystyle{ieeetr}
\bibliography{bibliography}